\documentstyle[twocolumn,epsf]{article}

\def\thebibliography#1{\section*{\normalsize \bf References 
 }\list
 {[\arabic{enumi}]}{\settowidth\labelwidth{[#1]}\leftmargin\labelwidth
 \advance\leftmargin\labelsep
 \usecounter{enumi}}
 \def\newblock{\hskip .11em plus .33em minus .07em}
 \sloppy\clubpenalty4000\widowpenalty4000
 \sfcode`\.=1000\relax}

\begin{document}

\twocolumn[

\begin{center} \large \bf 
   Interpolating self-energy of the infinite-dimensional
   Hubbard model: \\ Modifying the iterative perturbation theory
\end{center}
\vspace{-3mm}

\begin{center} 
   M. Potthoff, T. Wegner and W. Nolting
\end{center}
\vspace{-6mm}

\begin{center} \small \it 
   Lehrstuhl Festk\"orpertheorie,
   Institut f\"ur Physik, 
   Humboldt-Universit\"at zu Berlin, 
   D-10115 Berlin, 
   Germany
\end{center}
\vspace{2mm}

\begin{center}
\parbox{141mm}{ 
We develop an analytical expression for the self-energy of the
infinite-dimensional Hubbard model that is correct in a number
of different limits. The approach represents a
generalization of the iterative perturbation theory to arbitrary
fillings. In the weak-coupling regime perturbation theory to 
second order
in the interaction $U$ is recovered. 
The theory is exact
in the atomic limit. The high-energy behavior of the self-energy
up to order $1/E^2$ and thereby the
first four moments of the spectral density are reproduced correctly. 
Referring to a standard strong-coupling moment method, we analyze
the limit $U\mapsto \infty$. 
Different modifications of the approach are discussed and 
tested by comparing 
with the results of an exact diagonalization study.
}
\end{center}
\vspace{8mm} 
]

{\center \bf \noindent I. INTRODUCTION \\ \mbox{} \\} 

The theoretical understanding of correlated electron systems 
represents 
a central problem in condensed-matter physics. One of the simplest, 
but non-trivial models that describe interacting fermions on a 
lattice, is the Hubbard model \cite{Hub63,Gut63,Kan63}. It is 
nowadays 
studied intensively to gain insight into the fundamental mechanisms 
being responsible for itinerant magnetism, metal-insulator (Mott) 
transitions and high-temperature superconductivity. However, with 
the exception of the one-dimensional case \cite{LW68}, an exact 
solution is not known, and a completely satisfactory understanding 
of its properties has not yet been achieved.

As was shown by Metzner and Vollhardt \cite{MV89}, the Hubbard model
is simplified considerably in the limit of high spatial dimensions 
$d$. However, it still remains a non-trivial model, the 
essential properties of which are comparable to those at low
dimensions 
$d=2,3$. An exact solution of the $d=\infty$ Hubbard model or a 
reliable approximation for the entire range of the model parameters 
will thus provide a proper dynamical mean-field theory of the Hubbard 
model in any dimension \cite{Vol93,GKKR96}.

In the limit $d=\infty$ the Hubbard model can be mapped onto the 
single-impurity Anderson model (SIAM) supplemented by a 
self-consistency
condition \cite{GK92a,Jar92,JP93}. Thereby it becomes possible to
make use of various methods that are available for impurity problems.

Using the mapping onto an effective impurity problem, essentially 
exact results for the infinite-dimensional Hubbard model
can be obtained from quantum Monte Carlo (QMC) calculations
\cite{Jar92,RZK92,GK92b} 
and exact diagonalization methods (ED) 
\cite{CK94,SRKR94}. 
However, these approaches though being exact suffer from severe 
limitations.
Within ED calculations one is restricted to a small number of 
orbitals, and thus a smooth density of states cannot be obtained. 
On the other hand, QMC yields its results for the 
discrete Matsubara energies or along the imaginary time axis.
Therefore, it is difficult to access the low-temperature regime
where statistical errors become important within the QMC method. 
Furthermore, to obtain dynamical quantities such as the density of 
states, the analytical continuation to the real axis becomes 
necessary which cannot be performed reliably in all cases.
For these reasons the application of approximate methods for the 
infinite-dimensional Hubbard model still remains necessary. 

The development of approximate methods
should be guided by exactly solvable 
limiting cases and other rigorous analytical results available. 
Such information imposes a number of strong necessary conditions 
on any approximation. 
Moreover, comparison with the numerically exact results of ED or QMC 
techniques allows for a judgment of the quality of the
approximation a posteriori.

For the SIAM weak-coupling approaches are known to yield 
reliable quantitative information
\cite{YY70,YY75,Sal81,ZHS85}.
As has been demonstrated by Yosida and Yamada \cite{YY70,YY75}, 
perturbation theory in $U$ is quite well behaved for the symmetric 
case when expanding around the Hartree-Fock solution.

Based on these findings, Georges and Kotliar \cite{GK92a} introduced
a method called iterative perturbation theory (IPT) for the 
$d=\infty$ Hubbard model at half-filling ($n=1$). Within IPT the
SIAM is solved by means of second-order perturbation theory around
the Hartree-Fock solution (SOPT-HF) for arbitrary hybridization
functions followed by the self-consistent mapping onto the Hubbard
model. IPT leads to convincing results as has been proven by 
comparison with ED and QMC studies \cite{GK93,ZRK93,RKZ94}.
Furthermore, a direct treatment of zero temperature and real
energies is possible. The method turns out to be superior to the 
direct application of SOPT-HF to the Hubbard model. The latter does
not yield a metal-insulator transition, and Fermi-liquid behavior 
never breaks down \cite{WC94}.

The success of IPT may be partly due to the fact that at $n=1$ the
perturbational treatment accidentally reproduces the atomic limit. 
Away from half-filling this desirable property is lost, and
(the naive extension of) IPT is known to give unphysical results.

In a recent paper \cite{KK96} Kajueter and Kotliar proposed 
a modification of the IPT scheme introducing
an interpolating self-energy for the SIAM 
(see also: Refs.\ \cite{MRFBP82,MRLFT86}).
The self-energy exactly reproduces 
the trivially solvable limiting cases 
of vanishing interaction, $U=0$, of the completely filled and the 
empty band
(chemical potential $\mu \mapsto \pm \infty$), and the atomic limit. 
For small $U$ it is exact up to order $U^2$.
The low-energy behavior ($E\mapsto 0$) is fixed by imposing
the Friedel sum rule \cite{Lan66} (equivalently,
the Luttinger theorem \cite{LW60,MH89b}).
At half-filling the approach reduces to the usual IPT.
Finally, the self-energy has the correct asymptotic form for high
energies $E\mapsto \infty$: The first two coefficients within an
expansion in $1/E$ are reproduced exactly.
Results for the spectral density and the integrated spectral weight
based on the approach have been compared with ED calculations, and 
a rather convincing agreement between both methods has been 
observed indeed \cite{KK96}.

The correct high-energy behavior may be quite important, since this 
is closely connected with the moments of the spectral density 
\cite{Gor68}. The moments are defined by

\begin{equation}
  M^{(m)}_{d\sigma} = \frac{1}{\hbar} \int_{-\infty}^\infty E^m
  A_{d\sigma}(E) \, dE \: ,
\label{eq:momentsdef}
\end{equation}
where $A_{d\sigma}$ denotes the one-electron spectral density of 
the localized (d) state in the SIAM. The moments can be calculated
from the Hamiltonian directly \cite{NO81}. With increasing
$m$, however, they include equal-time correlation functions 
of higher and higher order. This fact limits the number of moments
that can be used in practice for the determination of the high-energy
behavior of the self-energy.

In their recent approach, Kajueter and Kotliar could take into account
exactly the moments up to $m=2$. The main purpose of this paper is to
demonstrate that improvement is still possible. Modifying their 
approach, we will show that a self-energy function can be 
constructed that respects the $m=3$ moment additionally, while
all other mentioned limiting cases are still recovered as before.
Thereby, higher-order correlation functions enter the theory. As
will be shown, these can be expressed without further approximations
in terms of the one-electron spectral density and can thus be
determined self-consistently.
Our analysis stresses the importance of this 
$m=3$ moment, especially what concerns spontaneous magnetism.

The interpolating self-energy for the SIAM is exact in the case of 
small $U$ 
and in the atomic limit; to see whether it can be considered as a 
reasonable interpolation between the weak- and the strong-coupling 
regime, we furthermore investigate (analytically) the limit 
$U\mapsto \infty$. Contact is made with a standard strong-coupling 
moment method [the spectral-density approach (SDA)] 
\cite{NB89,BdKNB90,PN96,HN96b}. 
The SDA, within the 
context of the Hubbard model, has proven its usefulness in several 
previous studies. Thereby, we can provide an additional
independent justification for the interpolating self-energy.

The theory is evaluated numerically. Different versions of the 
approach are tested, which are all compatible with the mentioned 
limiting cases. Finally, we compare with the results of an exact 
diagonalization study from Ref.\ \cite{KK96}.

{\center \bf \noindent II. MAPPING ONTO SIAM \\ \mbox{} \\} 

To begin with, we briefly recall the procedure by which the Hubbard
model can be mapped onto the SIAM and introduce some basic notations.

The Hubbard model reads:

\begin{equation}
  H = \sum_{ij\sigma} \left( T_{ij} - \mu \delta_{ij} \right)
  c^\dagger_{i\sigma} c_{j\sigma} 
  + \frac{1}{2} U \sum_{i \sigma} n_{i\sigma} n_{i -\sigma} \: .
\label{eq:hubbard}
\end{equation}
We consider a $d$-dimensional lattice with hopping between nearest
neighbors. 
Provided that the hopping integrals are scaled appropriately,
$T_{\langle ij \rangle}=t=t^\ast/(2\sqrt{d})$ 
($t^\ast=\mbox{const.}$),
a non-trivial model is obtained in the limit $d\mapsto \infty$
\cite{MV89}. The (non-interacting) Bloch-density of states for the 
$d=\infty$ simple cubic 
lattice is a Gaussian \cite{MV89}, a semi-elliptic
Bloch-density of states is obtained for the Bethe lattice 
with infinite coordination number \cite{vD91}.


The basic quantity to be calculated within the
model is the one-electron Green function

\begin{equation}
  G_{ij\sigma}(E) = \langle \langle c_{i\sigma} ; 
  c^\dagger_{j\sigma} \rangle \rangle_E \: .
\label{eq:green}
\end{equation}
Its diagonal elements can be written in the form

\begin{equation}
  G_\sigma(E) \equiv 
  G_{ii\sigma}(E) = \int_{-\infty}^\infty \frac{\hbar 
  \rho^{\rm (B)}(z)}
  {E-(z-\mu) - \Sigma_{\sigma}(E)} \: dz \: .
\label{eq:grhm}
\end{equation}
For $d\mapsto \infty$ the self-energy of the Hubbard model,
$\Sigma_{\sigma}(E)$, becomes $\bf k$ independent or site-diagonal
\cite{MV89,MH89b,Vol94}.

The Anderson model for a single impurity (SIAM) is given by:

\begin{eqnarray}
  H_{\rm SIAM} & = & \sum_{k\sigma} (\epsilon_k - \mu)
  c_{k\sigma}^\dagger c_{k\sigma} + \sum_\sigma (\epsilon_d - \mu)
  c_{d\sigma}^\dagger c_{d\sigma}
  \nonumber \\ & + &
  U n_{d\sigma} n_{d-\sigma} +
  \sum_{k\sigma} V_{kd} \left(
  c_{d\sigma}^\dagger c_{k\sigma} +
  c_{k\sigma}^\dagger c_{d\sigma}
  \right) \: . \nonumber \\
\end{eqnarray}
In all practical calculations the hybridization strength between the
conduction band and the localized d-level, $V_{kd}$, enters via the
the hybridization function which is defined by:

\begin{equation}
  \Delta(E) = \sum_k \frac{V_{kd}^2}{E-\epsilon_k} \: .
\label{eq:hybr}
\end{equation}
Let us also introduce the impurity Green function:

\begin{equation}
  G_{d\sigma}(E) = \langle \langle c_{d\sigma} ; c^\dagger_{d\sigma}
  \rangle \rangle_E \: .
\end{equation}
From its equation of motion we immediately have:

\begin{equation}
  G_{d\sigma}(E) = \frac{\hbar}{E - (\epsilon_d - \mu) - \Delta(E+\mu)
  - \Sigma_{d\sigma}(E)} \: ,
\label{eq:gds}
\end{equation}
where $\Sigma_{d\sigma}(E)$ is the d-level self-energy.

While in the $d=\infty$ limit of 
the Hubbard model all spatial degrees of 
freedom are frozen, the local (temporal) fluctuations still constitute
a non-trivial problem. This, however, is equivalent to the SIAM.
The Hubbard model and the SIAM can be connected by the
following self-consistency condition \cite{Jar92,JP93,GK92a}:

\begin{equation}
  \Delta_\sigma(E+\mu) = E - (\epsilon_d - \mu) - \Sigma_{\sigma}(E)
  - \hbar \left( G_{\sigma}(E) \right)^{-1} \: .
\label{eq:sc}
\end{equation}
$\Delta$ has to be interpreted as an effective hybridization function
which provides a coupling of the d level to the external bath of 
conduction electrons that simulates all temporal degrees of freedom 
in the Hubbard model. In the case of a ferromagnetic phase, the 
hybridization function must be spin-dependent. 
Provided that the condition (\ref{eq:sc}) is fulfilled, the 
self-energy 
of the Hubbard model is identical with the d-level self-energy of
the impurity problem, $\Sigma_{\sigma}(E) = \Sigma_{d\sigma}(E)$.
This also implies the corresponding identity between the respective
Green functions: $G_{\sigma}(E) = G_{d\sigma}(E)$.

If one is able to solve the SIAM for arbitrary hybridization functions
$\Delta_\sigma(E)$, the following two-step procedure for solving the
$d=\infty$ Hubbard model is suggested \cite{Jar92,JP93,GK92a}:
Given a hybridization function, we calculate (by solving the SIAM) the
self-energy $\Sigma_{d\sigma}(E)$ in the first step. In the second
step, using Eqs.\ (\ref{eq:grhm}) and (\ref{eq:sc}), a new
hybridization function is generated. Both steps are iterated  until
self-consistency is achieved.

In the following we concentrate on the first step which represents
the actual problem. We intend to derive an analytical expression for 
the self-energy of the SIAM that respects various exactly solvable
limiting cases and other rigorous facts available.

{\center \bf \noindent III. SPECTRAL MOMENTS \\ \mbox{} \\} 

The overall shape of the one-electron spectral density,

\begin{equation}
  A_{d\sigma}(E) = -\frac{1}{\pi} \mbox{Im} \, G_{d\sigma}(E+i0) 
  \: ,
\end{equation}
is fixed by their low-order spectral moments to a large extent. The
definition of the moments is given in Eq.\ (\ref{eq:momentsdef}). A
completely equivalent but independent representation is easily 
derived using the Heisenberg equation of motion for 
the time-dependent
operators in the definition of the spectral density. We obtain:

\begin{equation}
  M^{(m)}_{d\sigma} = \langle [ {\cal L}^m c_{d\sigma} , 
  c_{d\sigma}^\dagger ]_+ \rangle \: ,
\label{eq:moments}
\end{equation}
where ${\cal L O}=[{\cal O},H_{\rm SIAM}]_-$ 
denotes the commutator of an operator
$\cal O$ with the Hamiltonian, and $[\cdots , \cdots]_+$ is the
anticommutator. The straightforward calculation up to $m=3$ yields:

\begin{eqnarray}
  M^{(0)}_{d\sigma} & = & 1
  \nonumber \\
  M^{(1)}_{d\sigma} & = &
  \widetilde{\epsilon}_d + U \langle n_{d-\sigma} \rangle
  \nonumber \\
  M^{(2)}_{d\sigma} & = &
  \widetilde{\epsilon}_d^{\:2} + 2 \widetilde{\epsilon}_d U 
  \langle n_{d-\sigma} \rangle + U^2 \langle n_{d-\sigma} \rangle
  + \sum_k V_{kd}^2
  \nonumber \\
  M^{(3)}_{d\sigma} & = &
  \widetilde{\epsilon}_d^{\:3} + 3 \widetilde{\epsilon}_d^{\:2} U
  \langle n_{d-\sigma} \rangle
  \nonumber \\ & + & 
  \widetilde{\epsilon}_d U^2
  \langle n_{d-\sigma} \rangle ( 2 + \langle n_{d-\sigma} \rangle )
  + U^3 \langle n_{d-\sigma} \rangle
  \nonumber \\ & + &
  \sum_k V_{kd}^2 \left( \widetilde{\epsilon}_k 
  + 2 \widetilde{\epsilon}_d + 2 U \langle n_{d-\sigma} 
  \rangle \right)
  \nonumber \\ & + &
  U^2 \langle n_{d-\sigma} \rangle (1- \langle n_{d-\sigma} \rangle )
  \widetilde{B}_{d-\sigma} \: .
\end{eqnarray}
Here we have defined 
$\widetilde{\epsilon}_{d,k} = \epsilon_{d,k} - \mu$,
$\widetilde{B}_{d\sigma} = B_{d\sigma} - \mu$ and:

\begin{equation}
  \widetilde{B}_{d\sigma} = \widetilde{\epsilon}_d +
  \frac{1}{ \langle n_{d\sigma} \rangle 
  ( 1 - \langle n_{d\sigma} \rangle ) }
  \sum_k V_{kd} \langle c^\dagger_{k\sigma} c_{d\sigma} 
  (2 n_{d-\sigma} - 1) \rangle \: .
\label{eq:bdef}
\end{equation}
We notice that the $m=3$ moment includes higher-order correlation 
functions.

We can use these explicit results for the moments to fix the
high-energy behavior of the self-energy. For this purpose we consider
the following representation of the d Green function:

\begin{equation}
  G_{d\sigma}(E) = \int_{-\infty}^\infty \frac{A_{d\sigma}(E')}
  {E-E'} \: dE' \: .
\label{eq:gspden}
\end{equation}
Expanding the denominator in powers of $1/E$, we get:

\begin{equation}
  G_{d\sigma}(E) = \sum_{m=0}^\infty \frac{\hbar}{E^{m+1}} 
  M_{d\sigma}^{(m)} \: .
\label{eq:gdexp}
\end{equation}
The coefficients in the $1/E$ expansion of the self-energy,

\begin{equation}
  \Sigma_{d\sigma}(E) = \sum_{m=0}^\infty 
  \frac{1}{E^{m}} C^{(m)}_{d\sigma} \: ,
\label{eq:sigmaexp}
\end{equation}
can be obtained by inserting Eqs.\ (\ref{eq:gdexp}) and 
(\ref{eq:sigmaexp}) and the analogous expansion ot the
hybridization function into Eq.\ (\ref{eq:gds}):

\begin{eqnarray}
  C^{(0)}_{d\sigma} & = & U \langle n_{d-\sigma} \rangle
  \nonumber \\
  C^{(1)}_{d\sigma} & = & U^2 \langle n_{d-\sigma} \rangle
  \left( 1 - \langle n_{d-\sigma} \rangle \right)
\label{eq:sicoeff}
  \\
  C^{(2)}_{d\sigma} & = & U^2 \langle n_{d-\sigma} \rangle
  \left( 1 - \langle n_{d-\sigma} \rangle \right) \times
  \nonumber \\ && \hspace{-2mm}
  \left( \widetilde{B}_{d-\sigma} 
  + U (1- \langle n_{d-\sigma} \rangle) \right) \nonumber \: .
\end{eqnarray}
An approximate expression for the self-energy of the SIAM 
should be consistent with this rigorously
derived high-energy behavior.

The Hartree-Fock approximation for the self-energy,

\begin{equation}
  \Sigma_{d\sigma}^{(0)}(E) = U \langle n_{d-\sigma} \rangle \: ,
\end{equation}
only respects the zeroth-order coefficient in the high-energy
expansion. The zeroth and the first coefficient are reproduced by
the self-energy

\begin{equation}
  \Sigma_{d\sigma}^{(1)}(E) = U \langle n_{d-\sigma} \rangle
  +
  \frac{U^2 \langle n_{d-\sigma} \rangle 
  ( 1 - \langle n_{d-\sigma} \rangle) }
  {E+\mu-\epsilon_d-U(1-\langle n_{d-\sigma}
  \rangle)} \: ,
\label{eq:sighub}
\end{equation}
which is obtained when applying the Hartree-Fock decoupling scheme
not at the first but at the second level in the hierarchy of equations
of motion for the Green function. This is analogous to the 
``Hubbard-I'' approximation \cite{Hub63} 
within the context of the Hubbard model.
The simplest form of a self-energy that implies the correct expansion
coefficients up to order $1/E^2$ is given by:

\begin{equation}
  \Sigma_{d\sigma}^{(2)}(E) = U \langle n_{d-\sigma} \rangle
  +
  \frac{U^2 \langle n_{d-\sigma} \rangle 
  ( 1 - \langle n_{d-\sigma} \rangle) }
  {E+\mu-B_{d-\sigma}-U(1-\langle n_{d-\sigma}
  \rangle)} \: .
\label{eq:sigsda}
\end{equation}
The higher-order correlation functions included in $B_{d\sigma}$ have
to be determined self-consistently as well as the mean occupation
numbers $\langle n_{d\sigma} \rangle$. Both, $\Sigma^{(1)}$
and $\Sigma^{(2)}$, are correct in the atomic limit. $B_{d\sigma}$
reduces to $\epsilon_d$ in this case, and $\Sigma^{(1)}$ just 
coincides with the self-energy of the atomic limit. 

The self-energy (\ref{eq:sigsda}) is the result found within 
the spectral-density approach (SDA) \cite{NB89,BdKNB90,PN96,HN96b}. 
Actually, the SDA is a standard strong-coupling approach to the 
Hubbard model. Its main idea is to start from a two-pole ansatz
for the spectral density and to fix all parameters in the ansatz 
such that the moments (up to $m=3$) are correct. Recent 
investigations 
of the $d=2$ Hubbard model \cite{BE95,MEHM95} point out that the 
results of the SDA are identical to 
those of the Roth approach \cite{Rot69} and the Mori-Zwanzig 
projection technique \cite{Mor65a,Zwa61}. All methods yield a 
two-pole structure for the interacting Green function. Neglecting
quasi-particle damping (for damping effects cf.\ Ref.\ 
\cite{HN96a}), this
two-pole structure can be made plausible for the strong-coupling
regime: An analysis of Harris and Lange \cite{HL67} rigorously shows 
that additional structures merely have a spectral weight of the order 
$(t/U)^4$ or less. It is remarkable that such a simple form 
(\ref{eq:sigsda}) of the self-energy is able to reproduce at least
qualitatively the correct dispersion of the Hubbard bands. This has 
been 
proven for the $d=2$ case by comparisons \cite{BE95,MEHM95} with QMC 
\cite{BSW94a,BSW94b} and with ED calculations \cite{LLM+92} on small
square Hubbard arrays. We thus believe that the step going from the 
Hubbard-I solution, Eq.\ (\ref{eq:sighub}), to a solution, Eq.\ 
(\ref{eq:sigsda}), that respects the $m=3$ moment additionally is 
quite important, at least for $U\mapsto \infty$. It is even decisive 
if spontaneous magnetism is considered as it is known that 
Hubbard's original solution predicts magnetic order only under 
extreme circumstances \cite{Hub63}. Within the SDA the spin 
dependence of $B_\sigma$ induces a spin-dependent shift of the 
bands which favors magnetism. Indeed, the results for the ferro- 
and antiferromagnetic Hubbard model seem to be qualitatively 
correct \cite{magdisc}.
On the other hand, the SDA completely fails to reproduce Fermi-liquid 
behavior for small $U$. At half-filling it predicts an insulator 
for each $U>0$ and thereby is not able to describe the Mott 
transition as well. Apart from the neglection of quasi-particle
damping, however, the SDA yields plausible results for $U \mapsto
\infty$.

The conclusion from the preceding discussion should be that it may
be quite important to account for the $m=3$ moment, especially
what concerns spontaneous magnetism. 
This will guide our search for an approximate self-energy
in the case of the SIAM which can be identified with the 
self-energy of the $d=\infty$ Hubbard model via 
the self-consistent mapping.

{\center \bf \noindent IV. MODIFICATION OF IPT \\ \mbox{} \\} 

Our approach is a modification of the iterative perturbation theory
(IPT) \cite{GK92a}. Within the usual IPT the SIAM is solved by means
of perturbation theory up to second order in the coupling $U$. 
The self-energy is given by:

\begin{equation}
  \Sigma_{d\sigma}^{\rm (IPT)}(E) = U \langle n_{d-\sigma} \rangle + 
  \Sigma^{\rm (SOC)}_{d\sigma}(E) \: .
\label{eq:sigmaipt}
\end{equation}
The first-order term is the Hartree-Fock self-energy; the second-order
contribution reads:

\begin{eqnarray}
  && \hspace{-9mm} \Sigma^{\rm (SOC)}_{d\sigma}(E) = 
  \frac{U^2}{\hbar^3} 
  \! \int \!\!\! \int \!\!\! \int 
  \frac{A^{\rm (HF)}_{d\sigma}(x) A^{\rm (HF)}_{d-\sigma}(y) 
  A^{\rm (HF)}_{d-\sigma}(z)}{E-x+y-z} \times
  \nonumber \\ &&
  (f(x) f(-y) f(z) + f(-x) f(y) f(-z))
  \: dx \, dy \, dz \: .
  \nonumber \\ &&
\label{eq:sigma2}
\end{eqnarray}
Here $f(x)=1/(\exp(\beta x)+1)$ is the Fermi function, and 
$\beta=1/k_BT$. The Hartree-Fock spectral density

\begin{equation}
  A^{\rm (HF)}_{d\sigma}(E) = - \frac{1}{\pi} \mbox{Im} \, 
  G^{\rm (HF)}_{d\sigma}(E+i0)
\label{eq:rho1}
\end{equation}
is obtained from:

\begin{equation}
  G^{\rm (HF)}_{d\sigma}(E) = 
   \frac{\hbar} 
  {E - (\epsilon_d - \widetilde{\mu}_\sigma) 
  - \Delta_\sigma(E+\mu) - U \langle n_{d-\sigma} \rangle} \: .
\label{eq:g1}
\end{equation}
The parameter $\widetilde{\mu}_\sigma$ has been introduced for later 
purposes. Within IPT we have $\widetilde{\mu}_\sigma=\mu$.

Following Kajueter and Kotliar \cite{KK96}, we consider an ansatz 
for the self-energy:

\begin{equation}
  \Sigma_{d\sigma}(E) = U \langle n_{d-\sigma} \rangle +
  \frac{a_\sigma \Sigma^{\rm (SOC)}_{d\sigma}(E)}
  {1 - b_\sigma \Sigma^{\rm (SOC)}_{d\sigma}(E)} \: .
\label{eq:ansatz}
\end{equation}
$a_\sigma$, $b_\sigma$ and $\widetilde{\mu}_\sigma$ are treated as
free parameters which will be fixed such that the approximation 
becomes exact in a number of limiting cases. It is assumed that
(\ref{eq:ansatz}) provides a reasonable interpolation between the 
different limits.

In Ref.\ \cite{KK96} Kajueter and Kotliar determined the 
parameter $a_\sigma$ to get the correct $m=2$ moment of the 
resulting spectral density and the parameter $b_\sigma$ to get 
the correct result for the atomic limit. Here, in contrast 
$a_\sigma$ and $b_\sigma$ will be fitted 
to the $m=2$ and to the $m=3$ moments.

This can be performed straightforwardly: We start by expanding the 
denominator in Eq.\ (\ref{eq:sigma2}) in powers of
$1/E$ to obtain the high-energy behavior of 
$\Sigma^{\rm (SOC)}_{d\sigma}$:

\begin{equation}
  \Sigma_{d\sigma}^{\rm (SOC)}(E) = \sum_{m=1}^\infty 
  \frac{1}{E^{m}} D^{(m)}_{d\sigma} \: ,
\label{eq:sigma2exp}
\end{equation}
where the coefficients are given by:

\begin{eqnarray}
  D^{(1)}_{d\sigma} 
  & = & U^2 \langle n_{d-\sigma} \rangle^{\rm (HF)}
  \left( 1 - \langle n_{d-\sigma} \rangle^{\rm (HF)} \right)
  \label{eq:si2coeff}
  \nonumber \\
  D^{(2)}_{d\sigma} & = & U^2 \langle n_{d-\sigma} \rangle^{\rm (HF)}
  \left( 1 - \langle n_{d-\sigma} \rangle^{\rm (HF)} \right) \times
  \nonumber \\ && \hspace{-2mm}
  \left( {B}_{d-\sigma}^{\rm (HF)} - \widetilde{\mu}_\sigma
  + U \langle n_{d-\sigma} \rangle \right) \: .
\end{eqnarray}
Here

\begin{equation}
  \langle n_{d\sigma} \rangle^{\rm (HF)} =
  \frac{1}{\hbar}
  \int_{-\infty}^\infty f(E) A^{\rm (HF)}_{d\sigma}(E) \, dE
\label{eq:nhf}
\end{equation}
is a fictive (Hartree-Fock) particle number, and

\begin{eqnarray}
  && \hspace{-14mm} B_{d\sigma}^{\rm (HF)} = \epsilon_d + 
  \frac{1}{\langle n_{d\sigma} \rangle^{\rm (HF)} 
  ( 1 - \langle n_{d\sigma} \rangle^{\rm (HF)} )} \times
  \nonumber \\ && \hspace{2mm}
  \sum_{k} V_{kd}
  \langle c_{k\sigma}^\dagger c_{d\sigma} \rangle^{\rm (HF)}
  (2 \langle n_{d-\sigma} \rangle^{\rm (HF)} - 1) 
\label{eq:meanbandshift}
\end{eqnarray}
is the Hartree-Fock value of the higher-order correlation functions
defined in Eq.\ (\ref{eq:bdef}). Comparing with the exact coefficients
given in Eq.\ (\ref{eq:sicoeff}), we notice that the IPT
self-energy $\Sigma_{d\sigma}^{\rm (IPT)}$ does not have the 
correct high-energy behavior away from half-filling.

From the equation of motion for the Green function 
$\langle \langle c_{d\sigma} ; c^\dagger_{k\sigma} \rangle 
\rangle^{\rm (HF)}$ it can be seen
that the hybridization-induced correlation functions in the 
definition of 
$B_{d\sigma}^{\rm (HF)}$ can be expressed in terms of the 
localized HF Green function:

\begin{eqnarray}
  && \hspace{-10mm} \sum_k V_{kd} 
  \langle c_{k\sigma}^\dagger c_{d\sigma} \rangle^{\rm (HF)}  =
  -\frac{1}{\pi \hbar} \mbox{Im} \int_{-\infty}^\infty f(E) \times 
  \nonumber \\  && \hspace{5mm}
  \Delta_\sigma(E+i0+\mu) \:
  G^{\rm (HF)}_{d\sigma}(E+i0) \: dE \: .
\label{eq:nkdhf}
\end{eqnarray}

The high-energy behavior of the interpolating self-energy 
can be derived from the expansion (\ref{eq:sigma2exp}) and from 
(\ref{eq:ansatz}). Comparing with the exact coefficients of the
$1/E$ expansion in Eq.\ (\ref{eq:sicoeff}) again, we have to choose

\begin{equation}
  a_\sigma = \frac{\langle n_{d-\sigma} \rangle 
  (1 - \langle n_{d-\sigma} \rangle)}
  {\langle n_{d-\sigma} \rangle^{\rm (HF)} 
  ( 1 - \langle n_{d-\sigma} \rangle^{\rm (HF)} )}
\label{eq:apar}
\end{equation}
and 

\begin{equation}
  b_\sigma = \frac{ B_{d-\sigma} - \mu - B^{\rm (HF)}_{d-\sigma} 
  + \widetilde{\mu}_\sigma 
  + U (1 - 2 \langle n_{d-\sigma} \rangle ) }
  {U^2 \langle n_{d-\sigma} \rangle^{\rm (HF)} 
  ( 1 - \langle n_{d-\sigma} \rangle^{\rm (HF)} )}
\label{eq:bpar}
\end{equation}
to ensure the correct high-energy behavior of the self-energy 
$\Sigma_{d\sigma}$ and thereby the correct moments of the resulting
spectral density up to $m=3$. 
The result (\ref{eq:apar}) and (\ref{eq:bpar}) reduces to the 
approach of Ref.\ \cite{KK96} if $B_{d\sigma}$ and 
$B_{d\sigma}^{\rm (HF)}$ are replaced by $\epsilon_d$.

It is easily verified that our approach is correct in the atomic 
limit. Setting $V_{kd}=0$, the Hartree-Fock spectral density 
(\ref{eq:rho1}) reduces to a $\delta$ function which allows to
calculate the second-order contribution (\ref{eq:sigma2}) and thus
the self-energy (\ref{eq:ansatz}) immediately. It turns out to 
coincide with the self-energy of the atomic limit (\ref{eq:sighub}).

Next, we have to check the weak-coupling limit: Provided that the
parameter $\widetilde{\mu}_\sigma$ is chosen such that 
$\widetilde{\mu}_\sigma \mapsto \mu_0 \equiv \mu|_{U=0}$ 
for $U \mapsto 0$ (see below), we have $\langle n_{d\sigma} 
\rangle \mapsto \langle n_{d\sigma} \rangle^{\rm (HF)}$ for 
$U \mapsto 0$. Furthermore, since $B_{d\sigma} \mapsto 
B^{(\rm HF)}_{d\sigma}$ as $U\mapsto 0$, we have
$b_\sigma \sim 1/U$. Therefore, expanding the self-energy
(\ref{eq:ansatz}) in powers of $U$, we see that it is correct up to
order $U^2$ indeed. In particular, this implies that all Fermi-liquid 
properties as described in Ref.\ \cite{MH89b}
will be recovered for small $U$ at least.

Finally, the parameter $\widetilde{\mu}_\sigma$ has to be fixed. The
most natural choice is the following:

\begin{equation}
  \widetilde{\mu}_\sigma = \mu \: .
\label{eq:c1}
\end{equation}
Another possibility is due to an approach of Martin-Rodero et al.\ 
\cite{MRFBP82,MRLFT86} where $\widetilde{\mu}_\sigma$ is determined
from the condition:

\begin{equation}
  \langle n_{d\sigma} \rangle^{\rm (HF)} = 
  \langle n_{d\sigma} \rangle \: .
\label{eq:c2}
\end{equation}
In Ref.\ \cite{KK96} Kajueter and Kotliar imposed the Friedel sum rule
\cite{Lan66} as a condition to fix $\widetilde{\mu}_\sigma$.
Via the self-consistency condition (\ref{eq:sc}) this is equivalent
to the Luttinger theorem \cite{LW60} which in the case
of the $d=\infty$ Hubbard model reads \cite{MH89b}:

\begin{equation}
  \mu = \mu_0 + \Sigma_\sigma(0) \: .
\label{eq:c3}
\end{equation}

Let us briefly discuss the implications of the different choices.
First we notice that in all cases we have
$\widetilde{\mu}_\sigma \mapsto \mu_0$ for $U\mapsto 0$ as it must
be to ensure the correct weak-coupling behavior of the self-energy. 
Furthermore, the validity of the approach within all other limiting 
cases that have been considered is not affected by the condition 
chosen.

Inspecting the original derivation in Ref.\ \cite{LW60}, we recall 
that the validity of the Luttinger theorem depends on a number of 
presuppositions. For example, the theorem holds if perturbation
theory applies. Obviously, for small $U$ all conditions 
(\ref{eq:c1})--(\ref{eq:c3}) yield a theory that is compatible 
with the Luttinger theorem up to order $U^2$ at least. Here, 
another supposition is more important, namely
$\mbox{Im} \, \Sigma_\sigma(E) =0$ at $E=0$.
In particular, this implies $T=0$ \cite{tnonzero}. Therefore,
applying the third condition (\ref{eq:c3}) does not allow to 
consider finite temperatures. On the other hand, 
the conditions (\ref{eq:c1}) and (\ref{eq:c2}) do not suffer from 
this difficulty.

The second condition (\ref{eq:c2}) implies (for a constant 
hybridization function $\Delta_\sigma$) that 
$\widetilde{\mu}_\sigma=\mu_0+U\langle n_{d-\sigma} \rangle$.
This exactly compensates the energetic shift of the Hartree-Fock 
Green function (\ref{eq:g1}) by $U\langle n_{d-\sigma} \rangle$.
$G^{\rm (HF)}_{d\sigma}$ thereby becomes independent of $U$, and
$\Sigma^{\rm (SOC)}_{d\sigma}$ reduces to the second-order 
contribution from a weak-coupling theory in which the {\em free} 
($U=0$) instead of the HF d Green function is used in the calculation 
of the proper irreducible self-energy. This, however, has not to be 
confused with the plain (or conventional) weak-coupling theory, 
where the free chemical potential $\mu_0$ would have to be replaced 
by $\mu$ additionally and
which artificially breaks particle-hole symmetry
\cite{BJ90}. The second-order contribution within SOPT-HF is 
recovered only when using the first condition (\ref{eq:c1}).
Arguments in favor or against a particular weak-coupling approach 
have previously been developed by demanding the correct high-energy 
behavior (up to order $1/E$) \cite{GK92a,GK93}. Such reasoning,
however, is not meaningful in the present context since the correct 
high-energy behavior (up to order $1/E^2$) is reproduced in each 
case.

Let us mention that choosing $\widetilde{\mu}_\sigma=\mu$ (first 
condition) introduces a slight complication: Because of the shift of 
$G^{\rm (HF)}_{d\sigma}$ by the HF contribution 
$U\langle n_{d-\sigma} \rangle$, it may happen that
$\langle n_{d\sigma} \rangle^{\rm (HF)}=0$ (or
$\langle n_{d\sigma} \rangle^{\rm (HF)}=1$) for 
strong $U$ and $T=0$, which means that the parameters 
$a_{\sigma}$ and $b_\sigma$ are no longer well defined. 
However, within the limit 
$\langle n_{d-\sigma} \rangle^{\rm (HF)} \mapsto 0$ 
(or $\mapsto 1$) we have:

\begin{eqnarray}
  && \mbox{} \hspace{-14mm} 
  \frac{\Sigma^{\rm (SOC)}_{d\sigma}(E)}
  {\langle n_{d-\sigma} \rangle^{\rm (HF)}
  (1-\langle n_{d-\sigma} \rangle^{\rm (HF)})} 
  \longmapsto
  \nonumber \\ && \mbox{} \hspace{-7mm} 
  \frac{U^2}{\hbar^3} 
  \! \int \!\!\! \int \!\!\! \int 
  \frac{A^{\rm (HF)}_{d\sigma}(x) A^{\rm (HF)}_{d-\sigma}(y) 
  A^{\rm (HF)}_{d-\sigma}(z)}{E-x+y-z}
  \: dx \, dy \, dz \: .
\label{eq:sigman01}
\end{eqnarray}
Therefore, although $a_{\sigma}$ and $b_\sigma$ diverge, the 
interpolating self-energy remains finite.

So far it can be concluded that there are no differences between the 
three possibilities considered that are {\em crucial}. However, we 
notice that the third condition implies a restriction 
of the theory to zero temperature. 
We defer further discussion to Secs.\ VII and VIII.

{\center \bf \noindent V. SELF-CONSISTENT DETERMINATION OF 
HIGHER CORRELATION FUNCTIONS \\ \mbox{} \\} 

The definition of the parameter $b_\sigma$ involves the band-filling 
$\langle n_{d\sigma} \rangle$ and the higher-order correlation 
functions 
included in $B_{d\sigma}$ which are still unknown. A satisfactory 
theory cannot be constructed unless it is possible to determine these 
correlation functions without further approximations. No problems are
introduced for the band-filling which may be expressed in terms of the
spectral density:

\begin{equation}
  \langle n_{d\sigma} \rangle =
  \frac{1}{\hbar}
  \int_{-\infty}^\infty f(E) A_{d\sigma}(E) \, dE \: .
\label{eq:ndet}
\end{equation}

In the following we demonstrate that also $B_{d\sigma}$ can be reduced
to the spectral density or the Green function, respectively. According
to its definition, $B_{d\sigma}$ partly consists of a sum of 
one-particle
correlation functions. Applying the general spectral theorem 
\cite{FW71} and exploiting the equation of motion for the Green 
function
$\langle \langle c_{d\sigma} ; c^\dagger_{k\sigma} \rangle \rangle$,
we have:
\begin{eqnarray}
  && \hspace{-10mm} \sum_k V_{kd} 
  \langle c_{k\sigma}^\dagger c_{d\sigma} \rangle =
  -\frac{1}{\pi \hbar} \mbox{Im} \int_{-\infty+i0}^{\infty+i0} 
  f(E) \times 
  \nonumber \\  && \hspace{25mm}
  \Delta_\sigma(E+\mu) \:
  G_{d\sigma}(E) \: dE \: .
\label{eq:nkd}
\end{eqnarray}
Now we are left with the higher-order correlation functions only.
Using the commutator

\begin{equation}
  [ c_{d\sigma} , H ]_- = \widetilde{\epsilon}_d \, c_{d\sigma} +
  U c_{d\sigma} n_{d-\sigma} + \sum_p V_{pd} \, c_{p\sigma} \: ,
\label{eq:comm}
\end{equation}
the remaining terms in $B_{d\sigma}$ can be written in the form:

\begin{eqnarray}
  && \hspace{-12mm}
  \sum_k V_{kd}
  \langle c_{k\sigma}^\dagger c_{d\sigma} n_{d-\sigma} \rangle =
  - \, \frac{ \widetilde{\epsilon}_d }{U}
  \sum_k V_{kd} \, \langle c_{k\sigma}^\dagger c_{d\sigma} \rangle
  \nonumber \\ && \hspace{-12mm}
  - \frac{1}{U} \sum_{kp} V_{kd} V_{pd} 
  \langle c_{k\sigma}^\dagger c_{p\sigma} \rangle
  + \frac{1}{U} \sum_k V_{kd}
  \langle c_{k\sigma}^\dagger [ c_{d\sigma} , H ]_- \rangle \: .
\label{eq:hoc}
\end{eqnarray}
The first term on the right-hand side has just been treated above.
Using the spectral theorem and the equation of motion, one gets for 
the second one:

\begin{eqnarray}
  && \hspace{-5mm} \sum_{kp} V_{kd} V_{pd}
  \langle c_{k\sigma}^\dagger c_{p\sigma} \rangle =
  -\frac{1}{\pi \hbar} \mbox{Im} \int_{-\infty+i0}^{\infty+i0} 
  f(E) \times 
  \nonumber \\  && \hspace{-1mm}
  \Delta_\sigma(E+\mu) \: \left[ \:
  \Delta_\sigma(E+\mu) \, G_{d\sigma}(E) + \hbar \: \right]
  \: dE \: .
\label{eq:vvcorr}
\end{eqnarray}
Applying once more the general spectral theorem, performing a 
Fourier transformation to time representation and using the 
Heisenberg equation of motion, the third term can be written as:

\begin{eqnarray}
  && \hspace{-15mm}
  \sum_k V_{kd} 
  \langle c_{k\sigma}^\dagger [ c_{d\sigma} , H ]_- \rangle
  =
  \sum_k V_{kd}
  \left( - \frac{1}{\pi \hbar} \right) \mbox{Im}\times
  \nonumber \\ && \hspace{0mm}
  \int_{-\infty}^\infty \int_{-\infty}^\infty f(E)
  e^{iE(t-t')/\hbar} 
  \left(i \hbar \frac{\partial}{\partial t} \right) \times
  \nonumber \\ && \hspace{0mm}
  \langle \langle c_{d\sigma}(t) ; c_{k\sigma}^\dagger (t') 
  \rangle \rangle \: 
  d(t-t') \: dE \: .
\label{eq:hcorr}
\end{eqnarray}
Integration by part and back-transformation to energy representation
finally yields:

\begin{eqnarray}
  && \hspace{-10mm} \sum_k V_{kd} 
  \langle c_{k\sigma}^\dagger [ c_{d\sigma} , H ]_- \rangle \, = \,
  -\frac{1}{\pi \hbar} \mbox{Im} \int_{-\infty+i0}^{\infty+i0} 
  f(E) \times 
  \nonumber \\  && \hspace{21mm}
  E \: \Delta_\sigma(E+\mu) \,
  G_{d\sigma}(E) \: dE \: .
\label{eq:vvcorrfin}
\end{eqnarray}
Combining all results, we obtain:

\begin{eqnarray}
  \langle n_{d\sigma} \rangle ( 1 - \langle n_{d\sigma} \rangle ) \,
  \widetilde{B}_{d\sigma} & = & 
  \langle n_{d\sigma} \rangle ( 1 - \langle n_{d\sigma} \rangle ) \,
  \widetilde{\epsilon}_{d} \, - \, \nonumber \\
  && \hspace{-22mm}
  \frac{1}{\pi \hbar} \, \mbox{Im} 
  \int_{-\infty+i0}^{\infty+i0} f(E) \,
  \Delta_\sigma(E+\mu) \, \times
  \nonumber \\ && \hspace{-22mm}
  \left( \frac{2}{U} \Sigma_{d\sigma}(E)
  - 1 \right) \:
  G_{d\sigma}(E) \: dE \: ,
\label{eq:bfin}
\end{eqnarray}
where once more we exploited the equation of motion for $G_{d\sigma}$.
This completes the theory since $\langle n_\sigma \rangle$ and 
$B_{d\sigma}$ can be determined self-consistently from the d Green
function.

{\center \bf \noindent VI. STRONG-COUPLING LIMIT \\ \mbox{} \\} 

So far we have shown that the appropriate choice of the parameters
$a_\sigma$ and $b_\sigma$ in the ansatz (\ref{eq:ansatz}) yields a
self-energy that is exact in a number of limiting cases, namely 
trivially for $U=0$, for $\langle n_{d\sigma} \rangle=0$ and 
$\langle n_{d\sigma} \rangle=1$ and furthermore in the atomic limit 
and for small $U$ up to order $U^2$. The high-energy expansion
yields the correct coefficients up to order $(1/E)^2$. Thereby the 
first moments of the spectral density up to $m=3$ are reproduced 
exactly which is one additional moment compared with the approach
of Kajueter and Kotliar \cite{KK96}.

It is easily shown that the general particle-hole symmetry requires
$B_{d\sigma}$ to be zero at half-filling in the paramagnetic phase
(and for a symmetric Bloch-density of states). 
Since in this case $B_{d\sigma}^{\rm (HF)}=0$ and 
$\mu=\widetilde{\mu}_\sigma=U/2$, it follows that $b_\sigma=0$.
Consequently, the theory reduces to the conventional IPT for 
$\langle n_{d\sigma} \rangle = \langle n_{d-\sigma} \rangle = 0.5$.

We now turn over to the strong-coupling case $U\mapsto \infty$
in order to check whether the theory can be regarded as a reasonable
interpolation between the weak- and the strong-coupling regime. In 
the following we exclusively focus on the SIAM; any approach that 
uses the self-consistent mapping of the Hubbard model onto the SIAM 
can only be as reliable as the approximation employed for the solution
of the latter. Hence, for the present purposes the hybridization 
function $\Delta$ can be assumed to be independent of $U$.

Let us consider the first condition (\ref{eq:c1}) for determining
the fictive chemical potential, i.\ e.\ $\widetilde{\mu}_\sigma=\mu$.
As mentioned above, for small $U$ all Fermi-liquid properties are
recovered. For large $U$ the spectral density $A_{d\sigma}$ is 
expected to consist roughly of two dominant features separated by 
$U$ at about $\epsilon_d$ and $\epsilon_d + U$. Assuming 
$\langle n_{d\sigma} \rangle = \langle n_{d-\sigma} \rangle < 0.5$ 
for simplicity, the chemical 
potential will be located within the lower part: 
$\mu \sim \epsilon_d$.
Looking at the HF Green function (\ref{eq:g1}), we notice that 
$\langle n_{d-\sigma} \rangle^{\rm (HF)} = 0$ for $T=0$ and for $U$ 
larger than a certain critical value $U_{c1}$, and thus the relation 
(\ref{eq:sigman01}) applies. In particular, at the Fermi edge $E=0$: 

\begin{eqnarray}
  && \mbox{} \hspace{-14mm} 
  -\frac{1}{\pi} \, \mbox{Im} \: 
  \frac{\Sigma^{\rm (SOC)}_{d\sigma}(0)}
  {\langle n_{d-\sigma} \rangle^{\rm (HF)}
  (1-\langle n_{d-\sigma} \rangle^{\rm (HF)})} =
  \nonumber \\ && \mbox{} \hspace{-7mm} 
  \frac{U^2}{\hbar^3} 
  \! \int \!\!\! \int 
  A^{\rm (HF)}_{d\sigma}(x) A^{\rm (HF)}_{d-\sigma}(y) 
  A^{\rm (HF)}_{d-\sigma}(y-x)
  \: dx \, dy \: .
\label{eq:isigman01}
\end{eqnarray}
A simple analysis shows this two-fold convolution integral to be
non-zero for interactions $U$ smaller than another critical value
$U_{c2}$ ($>U_{c1}$). Via Eq.\ (\ref{eq:ansatz})
this implies a non-zero value of the imaginary part of the 
interpolating self-energy at $E=0$. For $U>U_{c2}$ we have
$\mbox{Im} \, \Sigma_{d\sigma}(E) \equiv 0$ in the vicinity of
$E=0$. Therefore, non-Fermi-liquid behavior is implied for all
$U>U_{c1}$.

On the contrary, at $T=0$ the second condition (\ref{eq:c2}) yields
$\mbox{Im} \, \Sigma_{d\sigma}(E) \sim E^2$ for $E\mapsto 0$ 
irrespective of the value for $U$.

Some more statements can be made if $U\mapsto \infty$. For this
purpose we first determine the total weight and the center of gravity 
of the imaginary part of the second-order contribution 
$\Sigma_{d\sigma}^{\rm (SOC)}$. Let us consider the integral
\begin{equation}
  \int_{-\infty}^\infty \: 
  E^m \left( - \frac{1}{\pi} \right) \mbox{Im}
  \Sigma_{d\sigma}^{\rm (SOC)}(E+i0) \, dE \: ,
\end{equation}
for $m=0,1$. Since $\Sigma_{d\sigma}^{\rm (SOC)}$ is analytical in
the upper half of the complex $E$-plane, the contour along the real 
axis can be deformed into a contour $C$ where each point of $C$ has 
a distance from $0$ that is larger than $R$ and $R\mapsto \infty$.
On the contour $C$ one is allowed to replace 
$\Sigma^{\rm (SOC)}_{d\sigma}$
by its asymptotic form (\ref{eq:sigma2exp}). After that $C$ can be
re-deformed into an integration along the real axis 
and the evaluation 
of the integral becomes trivial. The total weight and the center of 
gravity turn out to be $D_{d\sigma}^{(1)}$ and 
$D_{d\sigma}^{(2)}/D_{d\sigma}^{(1)}$, respectively, where
$D_{d\sigma}^{(m)}$ are the coefficients from the expansion 
(\ref{eq:sigma2exp}). Analogously, the total weight and the
center of gravity of $-\mbox{Im} \, \Sigma_{d\sigma}(E)/\pi$ are
given in terms of its $1/E$ expansion coefficients as
$C_{d\sigma}^{(1)}$ and
$C_{d\sigma}^{(2)}/C_{d\sigma}^{(1)}$.

At energies with 
$|E-D_{d\sigma}^{(2)}/D_{d\sigma}^{(1)}|\mapsto \infty$, 
the imaginary part of the second-order contribution vanishes, and 
the real part approaches:

\begin{equation}
  \mbox{Re} \: \Sigma_{d\sigma}^{\rm (SOC)}(E)
  \mapsto
  \frac{D_{d\sigma}^{(1)}}
  {E-D_{d\sigma}^{(2)}/D_{d\sigma}^{(1)}} \: .
\label{eq:sisocal}
\end{equation}
Now let us {\em assume} 
that there is a zero $E=E^{(0)}_\sigma$ of 
the denominator in Eq.\ (\ref{eq:ansatz}),

\begin{equation}
  1-b_\sigma \Sigma_{d\sigma}^{\rm (SOC)}(E^{(0)}_\sigma) = 0
  \: ,
\end{equation}
at a sufficiently high energy, i.\ e.\ we assume that

\begin{equation}
  |E^{(0)}_{\sigma} - D_{d\sigma}^{(2)}/D_{d\sigma}^{(1)}| 
  \mapsto \infty 
\label{eq:assu}
\end{equation}
for $U \mapsto \infty$.
This implies that the imaginary part of the interpolating 
self-energy
\begin{eqnarray}
  && \hspace{-8mm}
  -\frac{1}{\pi} \mbox{Im} \, \Sigma_{d\sigma}(E+i0) = \:
  \frac{a_\sigma}{\pi |b_\sigma|} \times
  \nonumber \\ && \hspace{-2mm}
  \frac{- |b_\sigma| \mbox{Im} \Sigma_{d\sigma}^{\rm (SOC)}(E)}
  {\left(1-b_\sigma \mbox{Re} \Sigma_{d\sigma}^{\rm (SOC)}(E) 
  \right)^2 + \left(-|b_\sigma| \mbox{Im} 
  \Sigma_{d\sigma}^{\rm (SOC)}(E) \right)^2 } 
  \nonumber \\ &&
\label{eq:imsigint}
\end{eqnarray}
has a $\delta$ singularity at $E=E^{(0)}_{\sigma}$. 
Using Eq.\ (\ref{eq:sisocal}) we can derive the asymptotic position

\begin{equation}
  E_\sigma^{(0)} \mapsto B_{d-\sigma} - \mu 
  + U (1-\langle n_{d-\sigma} \rangle)
\label{eq:epole}
\end{equation}
and the asymptotic weight of the $\delta$ function:
\begin{equation}
  -\frac{1}{\pi} \mbox{Im} \: \Sigma_{d\sigma}(E+i0) \mapsto U^2 
  \langle n_{d-\sigma} \rangle (1- \langle n_{d-\sigma} \rangle)
  \: \delta(E - E_\sigma^{(0)})
  \: .
\label{eq:imsig}
\end{equation}
The weight turns out to be equal to the {\em full} weight 
$C_{d\sigma}^{(1)}$ of $-\mbox{Im} \, \Sigma_{d\sigma}(E)/\pi$.
Subject to the assumption (\ref{eq:assu}), Eq. (\ref{eq:imsig}) tells
us that the weight of the $\delta$ peak at $E_\sigma^{(0)}$ will 
dominate $\mbox{Im} \, \Sigma_{d\sigma}(E)$ eventually. The real part 
can be obtained from a Kramers-Kronig-type relation:

\begin{equation}
  \mbox{Re} \: \Sigma_{d\sigma}(E) =
  U \langle n_{d-\sigma} \rangle \: - \: \frac{1}{\pi} 
  \: {\cal P} \!\!
  \int_{-\infty}^\infty 
  \frac{\mbox{Im} \, \Sigma_{d\sigma}(E'+i0)}{E-E'} \: dE' \: .
\label{eq:kk}
\end{equation}
It turns out that it is given by $\Sigma_{d\sigma}^{(2)}(E)$,
which is just the self-energy of the SDA 
[cf.\ Eq.\ (\ref{eq:sigsda})].

Eq.\ (\ref{eq:imsigint}) shows that the imaginary part of the
interpolating self-energy is non-vanishing at $E=E^{(0)}_\sigma$
as well as at those energies $E$ where 
$\mbox{Im} \, \Sigma_{d\sigma}^{\rm (SOC)}(E) \ne 0$, i.\ e.\ within
a certain energy interval around 
$D_{d\sigma}^{(2)}/D_{d\sigma}^{(1)}$.
We can thus consider $\Sigma_{d\sigma}(E)$ to consist of two 
additive parts. The dominating part of the self-energy has been 
identified as being equal to the self-energy of the SDA.
The remaining part does not vanish as $U\mapsto \infty$ 
as can be seen when expanding the interpolating self-energy 
(\ref{eq:ansatz}) in powers of $1/U$ directly. However, compared
with the SDA part, its weight is smaller by a factor $U^2$.
In fact, for strong $U$ the remaining part can be neglected 
completely, provided that its energetic position, which is well 
approximated by the center of gravity 
$D_{d\sigma}^{(2)}/D_{d\sigma}^{(1)}$ of 
$\Sigma_{d\sigma}^{\rm (SOC)}$, is well apart from $\epsilon_d$
and thereby insignificant with respect to the states that form
the lower Hubbard band (again we assume that 
$\langle n_{d\sigma} \rangle=\langle n_{d-\sigma} \rangle < 0.5$, 
for simplicity).

It can be concluded that for $U\mapsto \infty$ the 
interpolating self-energy $\Sigma_{d\sigma}(E)$ reduces to the 
SDA self-energy $\Sigma^{\rm (2)}_{d\sigma}(E)$, if two conditions 
are fulfilled: The first is given by (\ref{eq:assu}) and the second 
one reads: 

\begin{equation}
  | \, D_{d\sigma}^{(2)}/D_{d\sigma}^{(1)} - \epsilon_d \, | 
  \mapsto \infty 
\label{eq:assus}
\end{equation}
for $U\mapsto \infty$. Inserting Eq.\ (\ref{eq:epole}) and the 
coefficients from Eq.\ (\ref{eq:sigma2exp}) into 
(\ref{eq:assu}), the first condition can be rewritten:

\begin{eqnarray}
  && 
  U^2 \langle n_{d-\sigma} \rangle^{\rm (HF)}
  (1-\langle n_{d-\sigma} \rangle^{\rm (HF)}) \;
  |b_\sigma| \equiv 
  \nonumber \\ &&
  | B_{d-\sigma} - \mu - B_{d-\sigma}^{\rm (HF)}
  + \widetilde{\mu}_\sigma + U ( 1 -
  2 \langle n_{d-\sigma} \rangle) | \mapsto \infty
  \: .
  \nonumber \\ &&
\label{eq:sdacond}
\end{eqnarray}
As $U\mapsto \infty$ the correlation functions $B_{d\sigma}$ and
$B_{d\sigma}^{\rm (HF)}$ stay finite. Restricting ourselves to the
case $\langle n_{d\sigma} \rangle=\langle n_{d-\sigma} \rangle<0.5$, 
we have $\mu \mapsto \mbox{const.}$.
Thus we can write the two conditions in the form:

\begin{eqnarray}
  &&
  | \, \widetilde{\mu}_\sigma + 
  U (1-2\langle n_{d-\sigma} \rangle) \, | \mapsto \infty \: ,
  \label{eq:cc1} \\ &&
  | \, \widetilde{\mu}_\sigma - 
  U \langle n_{d-\sigma} \rangle \, | \mapsto \infty \: .
  \label{eq:cc2}
\label{eq:ccc}
\end{eqnarray}

In the following let us discuss the implications for the three 
choices that are considered for the determination of the parameter
$\widetilde{\mu}_\sigma$ according to Eqs.\ 
(\ref{eq:c1})-(\ref{eq:c3}). We start with the case 
$\widetilde{\mu}_\sigma=\mu$.
Obviously, both conditions (\ref{eq:cc1}) and (\ref{eq:cc2})
are fulfilled. Therefore, the first choice yields the SDA self-energy
in the limit $U\mapsto \infty$.

The second choice, $\langle n_{d\sigma} \rangle^{\rm (HF)} =
\langle n_{d\sigma} \rangle$, implies 
$\widetilde{\mu}_\sigma \sim U \langle n_{d-\sigma} \rangle$. 
While this fulfills the first 
condition, it is at variance with the second one. As 
$U\mapsto \infty$ the overall energy dependence of the 
self-energy is given by the SDA; there are, however, non-negligible 
modifications for energies around 
$E=D_{d\sigma}^{(2)}/D_{d\sigma}^{(1)}$, i.\ e.\ within the lower
Hubbard band. Compared with the SDA, the upper Hubbard band is 
completely unaffected, especially what concerns its spectral weight
and its center of gravity. This implies that also the weight and the
center of gravity of the lower Hubbard band agree with the
predictions of the SDA, since the (zeroth and the) first moment 
of the total spectral density is reproduced exactly. In contrast 
to the SDA, however, the non-zero imaginary 
part of the self-energy leads to quasi-particle damping 
within the lower band. Via the Kramers-Kronig-type relation
(\ref{eq:kk}) the quasi-particle energies will be modified, too.

The determination of $\widetilde{\mu}_\sigma$ according to the 
Friedel sum rule (or equivalently the Luttinger theorem) is more 
implicit. Nevertheless, the following indirect argument can be given:

We take $T=0$; furthermore we again restrict ourselves to
the case 
$\langle n_{d\sigma} \rangle=\langle n_{d-\sigma} \rangle<0.5$.
Let us first mention that for the general proof \cite{Lan66} of the 
Friedel sum rule for the SIAM one has to resort to various identities 
that apply to Fermi liquids. In particular, one needs \cite{Lan66}:

\begin{equation}
  \mbox{Im} \, \Sigma_{d\sigma}(E+i0) \sim E^2 
  \hspace{4mm} \mbox{for} \hspace{4mm} E \mapsto 0 \: .
\label{eq:imsigzero}
\end{equation}

Secondly, we show that the SDA is at variance with the sum rule
\cite{lutsda}. For $\langle n_{d\sigma} \rangle < 0.5$ the pole 
of the SDA self-energy at 
$E=B_{d-\sigma} - \mu - U (1-\langle n_{d-\sigma} \rangle)>0$
lies outside the range of integration. Thus 
$\Sigma_{d\sigma}^{(2)}(E)$ is real and 

\begin{equation}
  \frac{\partial \Sigma^{(2)}_{d\sigma}(E)}{\partial E}
  = \frac{-U^2 \langle n_{d-\sigma} \rangle 
  (1-\langle n_{d-\sigma} \rangle)}
  {(E+\mu-B_{d-\sigma}-U(1-\langle n_{d-\sigma} \rangle))^2}
  < 0 
\label{eq:dsigde}
\end{equation}
for all energies $-\infty < E < 0$. Since
$\mbox{Im} \, G_{d\sigma}(E+i0) \le 0$ for all $E$ and 
$\mbox{Im} \, G_{d\sigma}(E+i0) < 0$ for a certain 
energy range within $-\infty < E < 0$,
it follows that 

\begin{equation}
  \mbox{Im} \: \int_{-\infty+i0}^{i0} G_{d\sigma}(E)
  \frac{\partial \Sigma^{(2)}_{d\sigma}(E)}{\partial E}
  dE > 0 \: ,
\label{eq:fsrsda}
\end{equation}
which according to Ref.\ \cite{Lan66} implies that the Friedel sum 
rule is not obeyed.

Thirdly, since $\widetilde{\mu}_\sigma$ is fixed by imposing the 
validity of the Friedel sum rule and since the SDA implies
(\ref{eq:fsrsda}), we can conclude that
the interpolating self-energy must be different from the SDA 
self-energy in all cases, especially for $U\mapsto \infty$. 
Consequently, one of the two conditions (\ref{eq:cc1}) or 
(\ref{eq:cc2}) must be violated. If it is assumed that the 
first one holds, it follows that 
$\langle n_{d-\sigma} \rangle^{\rm (HF)}=0$ for $U$ larger than
a certain critical value. Analogously to the above discussion of 
the case $\widetilde{\mu}_\sigma=\mu$, this would imply
$\mbox{Im} \, \Sigma_{d\sigma}(0) \ne 0$, a consequence that is 
not compatible with the Friedel sum rule. Hence, we must have
$\langle n_{d-\sigma} \rangle^{\rm (HF)} \ne 0$ implying that
for $U\mapsto \infty$ the second condition does not hold. Similar
to the case $\langle n_{d\sigma} \rangle^{\rm (HF)}=
\langle n_{d\sigma} \rangle$, it can thus be concluded that apart from
quasi-particle damping the overall shape of the spectral density 
follows the predictions of the SDA, in particular, what concerns the 
energetic positions and the spectral weights of both Hubbard bands.

Summing up, it has turned out that all three choices to determine
$\widetilde{\mu}_\sigma$ more or less make contact with a standard
strong-coupling approach (SDA) for $U\mapsto \infty$. This fact 
provides additional justification for the interpolating self-energy
since the SDA is known to yield rather satisfactory results, at least 
on the qualitative level. Therefore, we believe 
that the theory is able 
to yield reliable results well beyond the weak-coupling regime.
Let us also mention that within this context it is important that the 
theory correctly accounts for the $m=3$ moment. Otherwise, we would
have ended up for $U\mapsto \infty$ with the ``Hubbard-I'' 
self-energy $\Sigma_{d\sigma}^{(1)}(E)$ only.

{\center \bf \noindent VII. RESULTS \\ \mbox{} \\} 

\begin{figure}[b] 
\unitlength1mm
\begin{picture}(0,58)
\put(0,0){\epsfxsize=70mm \epsfbox{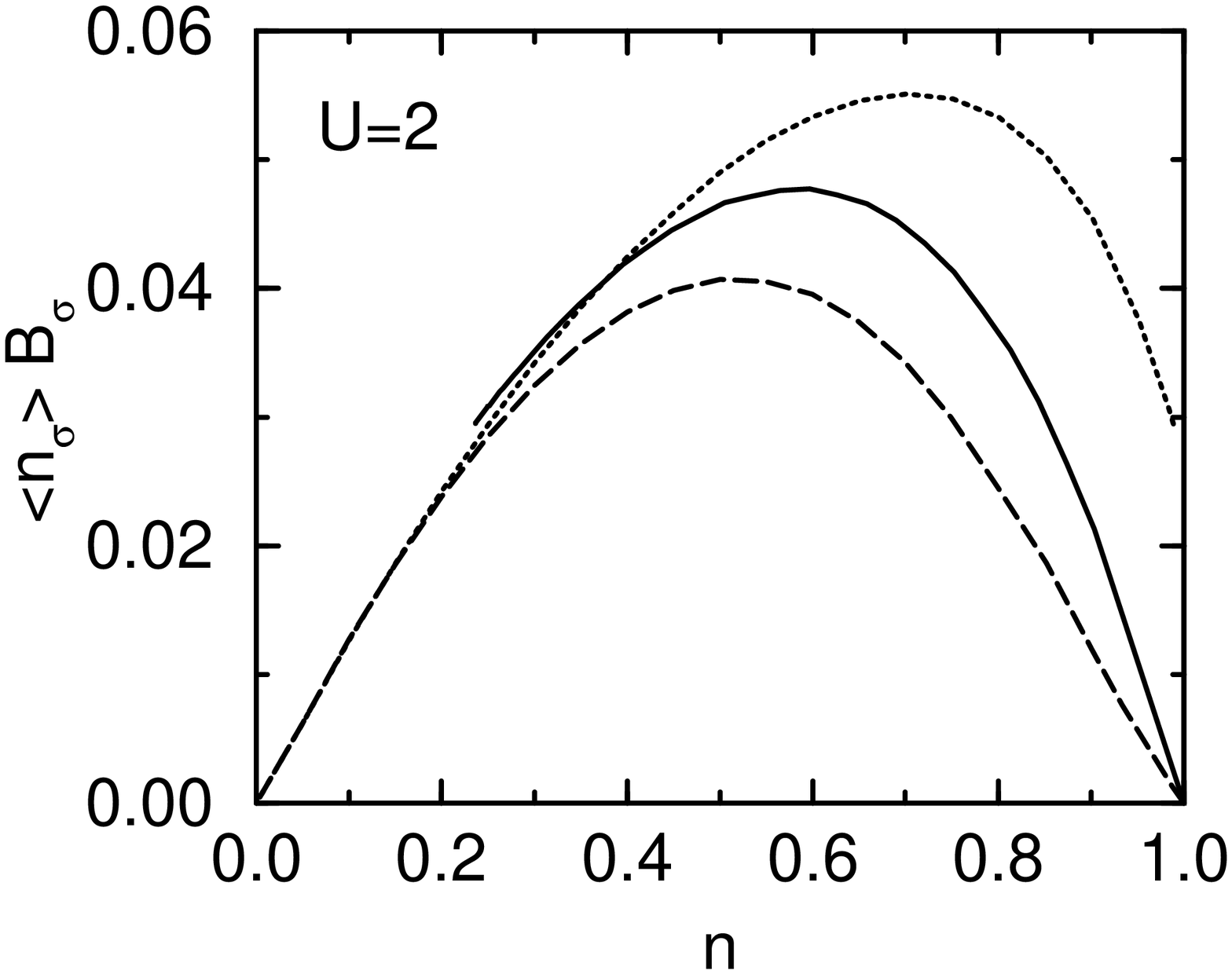}}
\end{picture}
\parbox[]{85mm}{\small Fig.~1.
``Band shift'' $\langle n_{\sigma} \rangle B_\sigma$ as a function 
of filling 
$n=\langle n_{\uparrow} \rangle + \langle n_\downarrow \rangle$
for $U=2$ (energies are given in units of $W$; $W$: width of the 
semi-elliptic Bloch-density of states).
Dotted line: calculation for $\widetilde{\mu}_\sigma = \mu$ [first
condition (\ref{eq:c1})].
Dashed line: calculation with $\widetilde{\mu}_\sigma$ being
determined by
$\langle n_{\sigma} \rangle^{\rm (HF)} = \langle n_{\sigma} \rangle$
[second condition (\ref{eq:c2})].
Solid line: $\mu=\mu_0+\Sigma_\sigma(0)$
[third condition (\ref{eq:c3})].
}
\end{figure}

We have evaluated the theory numerically. The procedure is described
briefly in Ref.\ \cite{KK96}. The additional computational effort due 
to the inclusion of the higher-order correlation functions via 
$B_{d\sigma}$ is almost negligible, and thus the algorithm remains
comparatively fast. The results being discussed in the following have 
been obtained for the Bethe lattice with infinite coordination number.
The semi-elliptic Bloch-density of states has a finite width $W$. All 
energies are given in units of $W$. Furthermore, we choose
$\epsilon_d=0$.

Within the SDA the correlation functions $B_{d\sigma}$ lead to an
additional energetic shift of the lower and the upper Hubbard band.
For strong $U$ the effective shift of the lower 
$-\sigma$ Hubbard band
is given by $\langle n_{\sigma} \rangle \, B_{\sigma}$ \cite{GN88}. 
Fig.~1 shows the dependence of this ``band shift'' on the occupation
number $n$ for $U=2$ as obtained from our modified IPT. Results for 
the three different conditions (\ref{eq:c1})--(\ref{eq:c3}) are shown.
In all cases we find a non-zero but small band shift. Except for the
case $\widetilde{\mu}_\sigma=\mu$, the curves closely resemble the 
corresponding results of the SDA \cite{GN88} for a $d=3$ bcc lattice.
This regards the absolute magnitude as well as the overall dependence 
on $n$. The calculation for $\widetilde{\mu}_\sigma=\mu$ yields a
non-zero value for $B_\sigma$ as $n \mapsto 1$, while it results in
$B_\sigma=0$ for $n=1$. Contrary, the second and the third condition
predict a continuous dependence on $n$ at half-filling. Another 
difficulty is observed for the case $\mu = \mu_0 + \Sigma_\sigma(0)$.
Below $n=0.23$ a self-consistent solution could not be found. This
also holds true (at a slightly different $n$) if we set $B_\sigma=0$.

\begin{figure}[b] 
\unitlength1mm
\begin{picture}(0,60)
\put(0,0){\epsfxsize=70mm \epsfbox{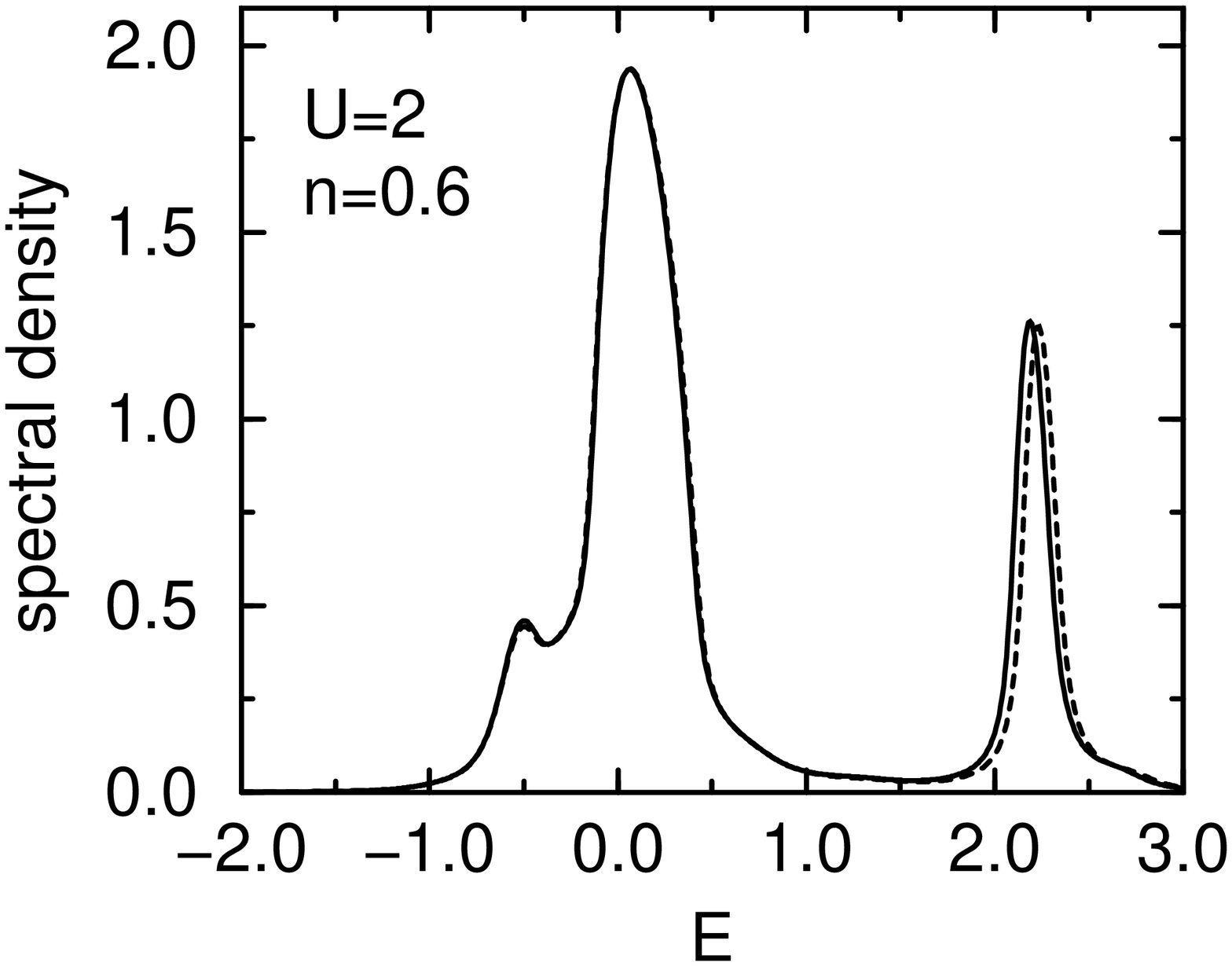}}
\end{picture}
\parbox[]{85mm}{\small Fig.~2.
Spectral density as a function of energy for $U=2$ and $n=0.6$.
Calculations assuming $\mu=\mu_0+\Sigma_\sigma(0)$.
Solid line: complete theory. Dashed line: $B_{\sigma} = 0$.
}
\end{figure}

The effect of $B_\sigma$ on the spectral density can be seen in 
Fig.~2 where we compare the result for the complete theory with 
the result for $B_\sigma=0$ (the approach of Ref.\ \cite{KK96}).
In both cases we have chosen the third condition (\ref{eq:c3}) to
determine $\widetilde{\mu}_\sigma$. The occupation number has been 
fixed at $n=0.6$ where according to Fig.~1 the band shift is at its
maximum. Qualitatively similar to the spectra expected for the SIAM,
both curves clearly show up three features: the lower and the upper
Hubbard band at $E\approx -0.5$ and $E \approx 2.2$ and a peak around
$E=0$, which is a reminiscence of the Kondo resonance being strongly 
broadened away from half-filling. We notice that the difference 
between both spectra is rather small. While the low-energy features 
are completely unaffected, the upper Hubbard band slightly shifts to 
lower energies when taking into account the $m=3$ moment. This is 
contrary to the SDA which predicts an energetic shift of the upper
Hubbard band (with respect to the Fermi energy) to higher energies 
by an amount $( 1 - n ) B_\sigma>0$. The effect can be traced back
to the (implicit) $U$ dependence of the hybridization function.

It has not yet been finally clarified what is the optimum choice to 
determine the fictive chemical potential $\widetilde{\mu}_\sigma$.
For this purpose we compare with results 
from the exact diagonalization
method of Caffarel and Krauth \cite{CK94,SRKR94}. We take the data
from Kajueter and Kotliar \cite{KK96} for 8 sites, $U=2$ and $n=0.86$.
Because of the finite number of orbitals considered 
in the calculation, 
the resulting spectral density is not smooth. Rather than comparing 
the spectral densities directly, a comparison of the integrated 
spectral weight is more appropriate. This is shown in Fig.~3.
The agreement of the ED result with the modified IPT calculation is
well provided that condition (\ref{eq:c2}) or (\ref{eq:c3}) is used 
(curves b) and c)). In both cases the residual discrepancies can be 
attributed to the finite system size 
in the ED calculation. Significant 
differences, however, are observed between the ED result and the 
calculation for $\widetilde{\mu}_\sigma=\mu$ (curve a)). 
In particular,
spectral weight is missing around $E=-0.5$.

\begin{figure}[t] 
\unitlength1mm
\begin{picture}(0,58)
\put(0,0){\epsfxsize=70mm \epsfbox{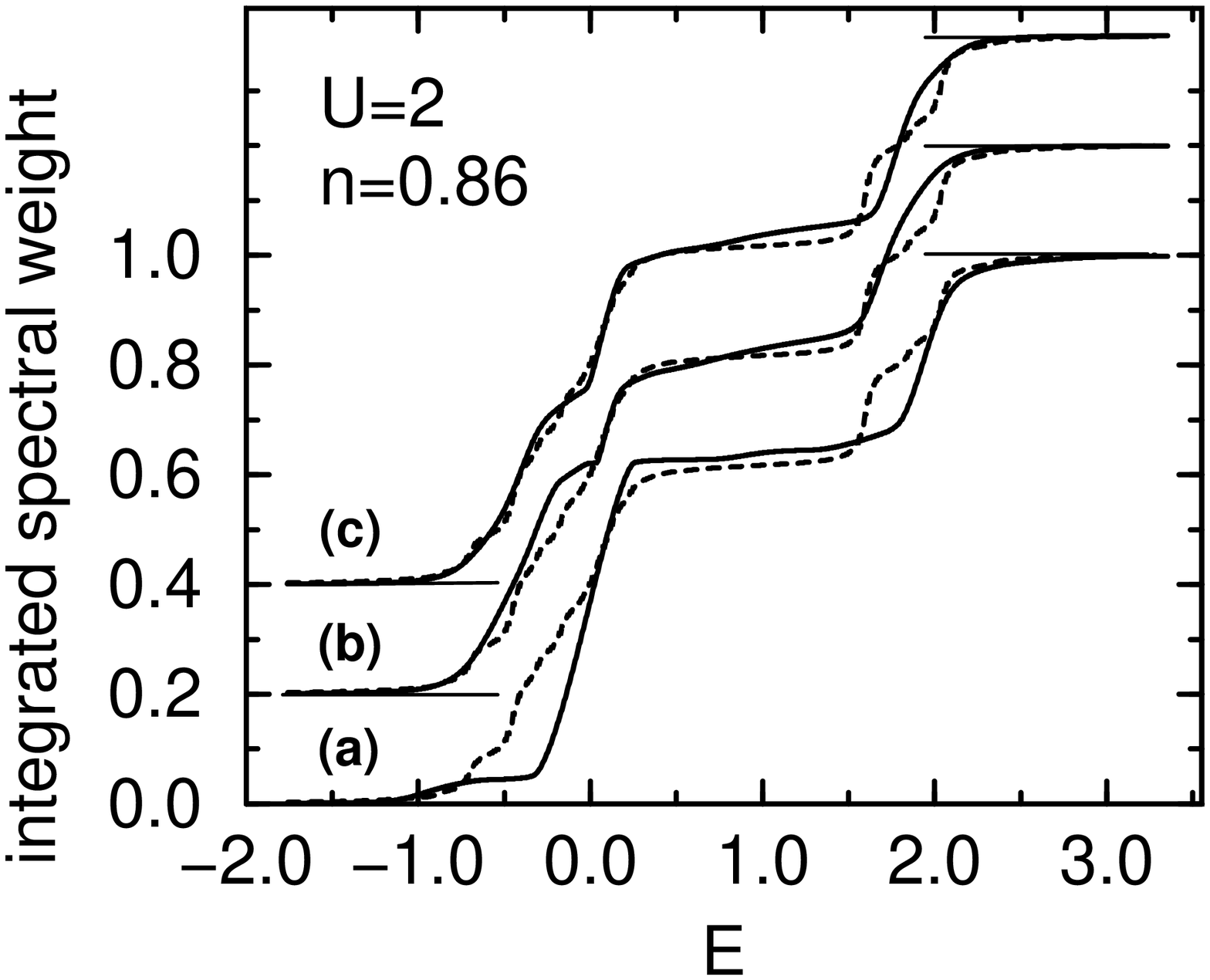}}
\end{picture}
\parbox[]{85mm}{\small Fig.~3.
Integrated spectral weight as a function of energy for $U=2$ and 
$n=0.86$. Solid lines: result for
(a) $\widetilde{\mu}_\sigma = \mu$.
(b) $\langle n_{\sigma}\rangle^{\rm (HF)}=\langle n_{\sigma}\rangle$,
(c) $\mu=\mu_0+\Sigma_\sigma(0)$,
Dashed line (a-c): exact diagonalization (8 sites), 
from Ref.\ \cite{KK96}, slightly smoothed.
(The vertical scale applies to (a). The curves (b) and (c) have been 
shifted constantly.)
}
\end{figure}

\begin{figure}[b] 
\unitlength1mm
\begin{picture}(0,58)
\put(0,0){\epsfxsize=70mm \epsfbox{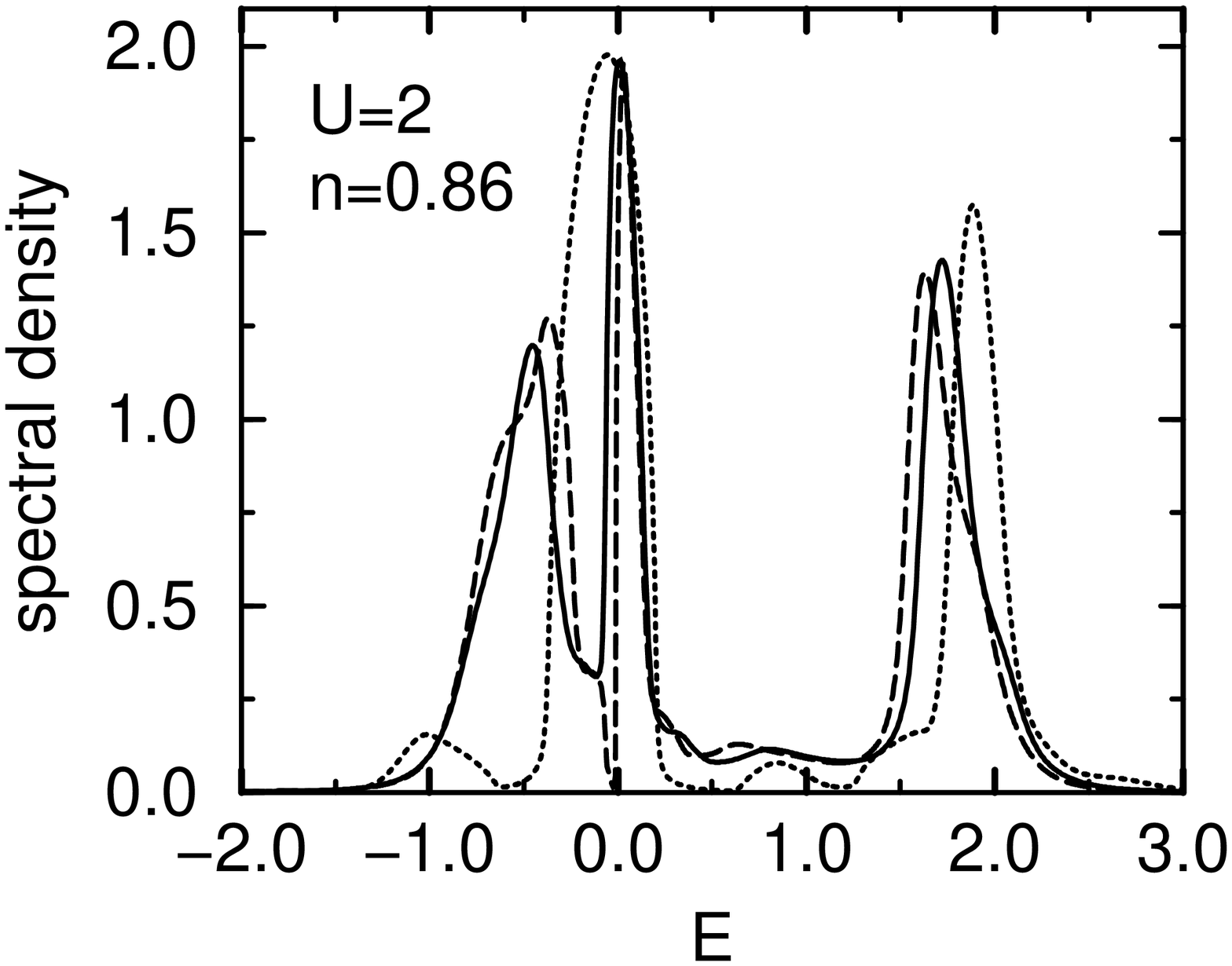}}
\end{picture}
\parbox[]{85mm}{\small Fig.~4.
Spectral density for $U=2$ and $n=0.86$.
Dotted line: $\widetilde{\mu}_\sigma = \mu$.
Dashed line: $\langle n_{\sigma} \rangle^{\rm (HF)} = 
\langle n_{\sigma} \rangle$. 
Solid line: $\mu=\mu_0+\Sigma_\sigma(0)$.
}
\end{figure}

Fig.~4 shows the corresponding spectral densities. We notice that
there are only minor differences between the results for 
$\langle n_{\sigma} \rangle^{\rm (HF)} = \langle n_{\sigma} \rangle$
and $\mu=\mu_0+\Sigma_\sigma(0)$. Apart from the lower and the upper
Hubbard band the spectra exhibit a sharp (Kondo) resonance at $E=0$.
On the other hand, the spectral density that is calculated for 
$\widetilde{\mu}_\sigma = \mu$ looks completely different. One can 
no longer distinguish unambiguously between the lower Hubbard band 
and
the resonance. At $E=-0.6$ a minimum can be found. The corresponding 
nearly constant trend of the integrated weight in Fig.~3 at the same 
energy, however, is at variance with the ED result which predicts a 
steep increase. Furthermore, the upper Hubbard band is significantly 
shifted to higher energies compared with the results for the second 
and the third condition which according to Fig.~3 reliably reproduce 
the peak position.

\begin{figure}[t] 
\unitlength1mm
\begin{picture}(0,58)
\put(0,0){\epsfxsize=70mm \epsfbox{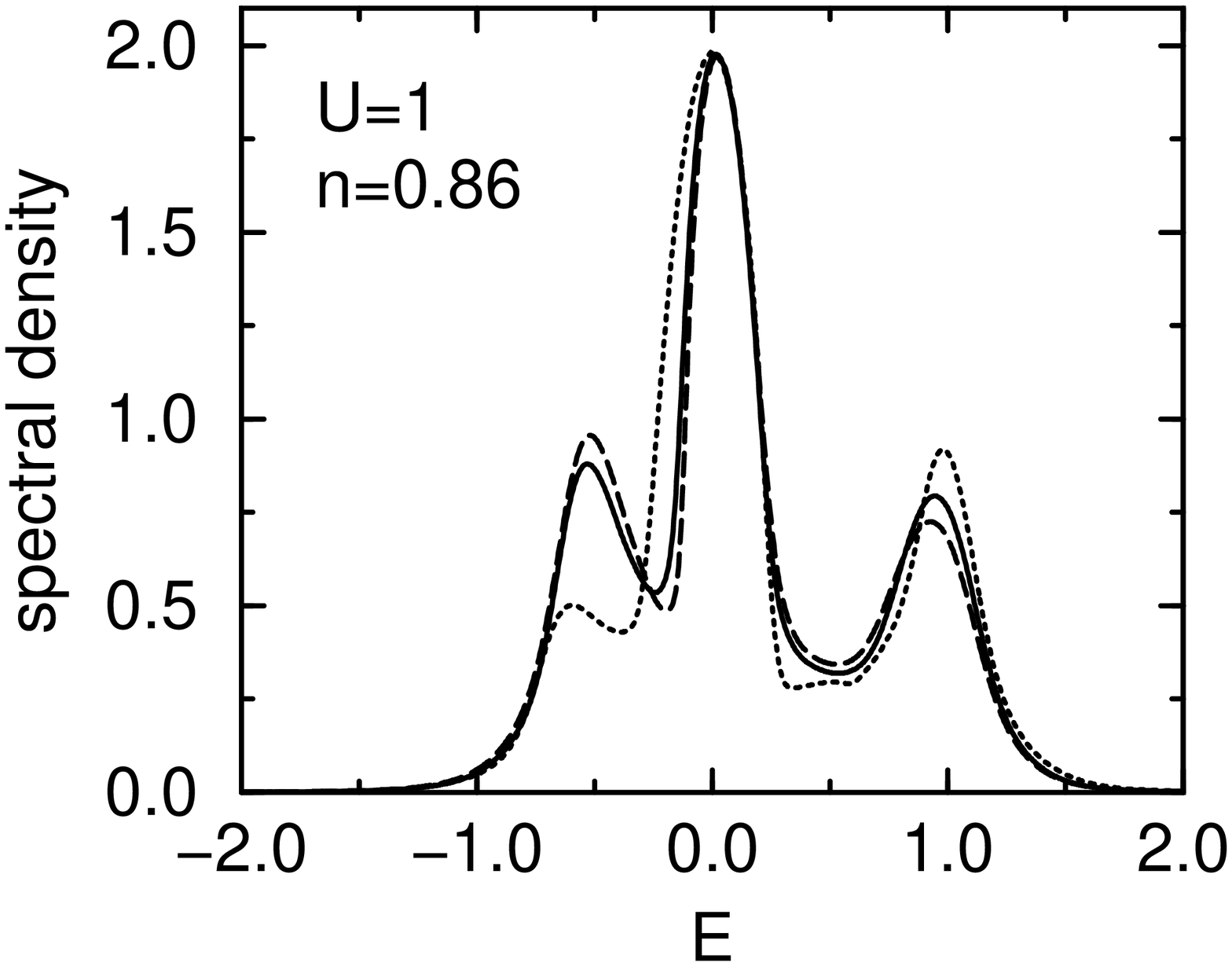}}
\end{picture}
\parbox[]{85mm}{\small Fig.~5.
The same as Fig.~4, but $U=1$.
}
\end{figure}

There is one unsatisfactory property of the modified IPT when using
the second condition 
$\langle n_{\sigma} \rangle^{\rm (HF)} = \langle n_{\sigma} \rangle$,
namely the drop of the spectral density just below the Fermi edge 
$E=0$ as can be seen in Fig.~4. This behavior, however, is only 
found for very strong interaction. Fig.~3 shows results for $U=1$
which still means strong correlation. Here we notice an almost perfect
agreement between the results for the second and the third condition.
Although much closer agreement between all three versions is obtained
generally if the interaction is reduced from $U=2$ to $U=1$, 
the result
for $\widetilde{\mu}_\sigma = \mu$ is still significantly different.
Remarkably, all three curves in Fig.~5 almost coincide at $E=0$.
According to Ref.\ \cite{MH89b} this indicates that Luttinger's
theorem is fulfilled approximately when using the first (\ref{eq:c1})
or second (\ref{eq:c2}) condition.

{\center \bf \noindent VIII. CONCLUSIONS AND OUTLOOK \\ \mbox{} \\} 

In this paper we have presented a modification of the recent approach 
of Kajueter and Kotliar \cite{KK96}. Using the self-consistent
mapping onto the SIAM, an approximate analytical expression for
the self-energy of the infinite-dimensional Hubbard model could
be constructed that reproduces a number of exactly solvable limits.
The conceptual improvement consists in the consideration of an
additional, the $m=3$, moment of the spectral density. It has been
shown that the higher-order correlation functions that are included
in the extra term $B_\sigma$ can be expressed without further
approximations by means of the spectral density. This allows for
a self-consistent (numerical) solution. The additional computational 
effort needed is almost negligible. The Green function on the real 
axis at $T=0$ can be computed fast compared with QMC or ED techniques.

The theory contains a fictive chemical potential 
$\widetilde{\mu}_\sigma$ which is considered as a free parameter that
can be fixed by a rather arbitrary condition without losing rigor
in all limiting cases mentioned. In this paper we have taken into
account three different possibilities to determine
$\widetilde{\mu}_\sigma$. The numerical results prove that the
different choices may imply considerable differences between the 
shapes of the resulting spectral densities, especially for very
strong interaction $U$. Thus further information was needed to get
a conclusive theory. We have compared our results with the data of an
ED study taken from Ref.\ \cite{KK96}. By the comparison the most
simple choice $\widetilde{\mu}_\sigma=\mu$ is excluded.

For both choices, $\mu = \mu_0 + \Sigma_\sigma(0)$ and 
$\langle n_\sigma\rangle^{\rm (HF)} = \langle n_\sigma \rangle$,
we encountered a minor difficulty: self-consistent solutions could
not be found for fillings below $n=0.23$ (at $U=2$) when taking the
first one; using the latter, we observed an implausible drop of the
spectral density just below the Fermi edge which, however, is present
in the case of very strong interaction only.

The condition $\mu = \mu_0 + \Sigma_\sigma(0)$ represents the 
Luttinger theorem for the $d=\infty$ Hubbard model. Imposing the
Luttinger theorem as a condition to fix $\widetilde{\mu}_\sigma$
as has been suggested by Kajueter and Kotliar \cite{KK96} implies
a considerable restriction of the theory: in this form the theorem
is only meaningful for a paramagnet at $T=0$ 
\cite{tnonzero}. This disadvantage is not present when using the 
condition
$\langle n_\sigma\rangle^{\rm (HF)} = \langle n_\sigma \rangle$
which was introduced by Martin-Rodero et al.\ originally 
\cite{MRFBP82,MRLFT86}. Finite temperatures and ferro- or 
antiferromagnetism can be treated without difficulties. 
Furthermore, the condition is much easier to handle numerically.
On the other hand, the difference found between the numerical results 
for the spectral density using either $\mu = \mu_0 + \Sigma_\sigma(0)$
or $\langle n_\sigma\rangle^{\rm (HF)} = \langle n_\sigma \rangle$ are
rather small, and the agreement with the ED data is equally good.

The usefulness of the $m=3$ moment is apparent in the limit of strong
correlations $U\mapsto \infty$. This limit of the approach has been 
investigated within the SIAM, i.\ e.\ for a fixed hybridization 
function $\Delta(E)$. The mean energetic positions and the weights
of the upper and the lower Hubbard band agree with the predictions 
of the SDA and with the exact results of Harris and Lange \cite{HL67}.
Here the $m=3$ moment turns out to be decisive. Otherwise, one would 
have ended up with the ``Hubbard-I'' solution only. 

The results for the paramagnetic $d=\infty$ Hubbard model on the 
Bethe lattice at $T=0$ have shown the effect of $B_\sigma$ on the 
spectral density to be rather small. Previous studies, however, 
strongly suggest that the $m=3$ moment is quite important in the 
context of spontaneous magnetism. This is obvious, for instance, 
when comparing the SDA (correct moments up to $m=3$) with the 
Hubbard-I solution (correct moments up to $m=2$). While the Hubbard-I 
solution yields magnetic order only under extreme circumstances, 
magnetism is favored within the SDA: The term $B_\sigma$ opens 
the possibility for a spin-dependent band shift. Consistent with 
the results found here, the effect of $B_\sigma$ in the paramagnetic 
phase is small within the SDA as well 
\cite{GN88}. Comparing the Hubbard-III 
alloy-analogy solution with a recently developed modification 
\cite{HN96a} where again $B_\sigma$ is included additionally, also 
stresses the importance of the $m=3$ moment for spontaneous magnetism.

The application of the presented method to magnetic phases represents
an interesting task for future studies. Let us mention that 
ferromagnetism in the $d=\infty$ Hubbard model for an fcc-type lattice
has been found recently in a QMC calculation \cite{Ulm96}.
Particle-hole symmetry requires $B_\sigma=0$ at half-filling (for
a symmetric Bloch-density of states) in the paramagnetic phase. In 
this case the usual IPT is recovered. However,
$B_\sigma \ne 0$ is possible for an antiferromagnet at half-filling. 
Future work may thus check whether the approach can improve the IPT
results for antiferromagnetic order at $n=1$ which are not completely 
satisfactory \cite{RKZ94}.

{\center \bf \noindent ACKNOWLEDGEMENT \\ \mbox{} \\} 

Support of this work by the Deutsche Forschungsgemeinschaft within
the Sonderforschungsbereich 290 is gratefully acknowledged.


\begin{thebibliography}{10}
\bibitem{lutsda}
  Within the context of the Hubbard model, Beenen and Edwards 
  \cite{BE95} pointed out that the Roth method, equivalently the
  SDA, is at variance with Luttinger's theorem for $U\mapsto \infty$.
\end{thebibliography}
\end{document}